\documentclass[useAMS]{mn2e}
\usepackage{graphicx}
\usepackage{epsfig}
\usepackage{amssymb}
\usepackage{lscape}
\usepackage{ulem}
\usepackage{txfonts}

\def\kms{km~s$^{-1}$}

\def\lya{Ly$\alpha$}

\def\HI{H\,{\sc i}}

\def\c2s{C\,{\sc ii}$^{\star}$}

\def\LIR{L$_{\rm IR}$}

\def\fgas{$f_{\rm gas}$}
\def\dfgas{$\Delta f_{\rm gas}$}
\def\dfgash2{$\Delta f_{\rm gas, H_2}$}
\def\mhi{M$_{\rm HI}$}

\title[HI in AGN hosts] {Atomic gas fractions in active galactic nucleus host galaxies}

\author[Ellison et al.] {Sara L. Ellison$^1$, Toby Brown$^2$, Barbara Catinella$^{3,4}$, Luca Cortese$^{3,4}$ \\ 
$^1$ Department of Physics \& Astronomy, University of Victoria, Finnerty Road, Victoria, British Columbia, 
  V8P 1A1, Canada\\
$^2$ Department of Physics \& Astronomy, McMaster University, 1280 Main St. W., Hamilton, Ontario L8S 4M1, Canada\\  
$^3$ International Centre for Radio Astronomy Research, The University of Western Australia, 35 Stirling Hwy, 
Crawley, WA 6009, Australia\\
$^4$ ARC Centre of Excellence for All Sky Astrophysics in 3 Dimensions (ASTRO 3D)
}
\begin{document}

\maketitle

\begin{abstract}

  The feedback from an active galactic nucleus (AGN) is frequently invoked
  as a mechanism through which gas can be heated or removed from a galaxy.
  However, gas fraction measurements in AGN hosts have yielded mixed support for
  this scenario.  Here, we re-visit 
  the assessment of \fgas\ (=M$_{HI}$/M$_{\star}$) in z$<$0.05 AGN hosts in the Sloan
  Digital Sky Survey (SDSS) using two complementary techniques.  First, we investigate
  \fgas\ for 75 AGN host galaxies in the 
  extended GALEX Arecibo SDSS Survey (xGASS), whose atomic gas fractions are complete
  to a few percent.  Second, we construct \HI\ spectral stacks
  of 1562 AGN from the Arecibo Legacy Fast ALFA (ALFALFA) survey, which enables us
  to extend the AGN sample
  to lower stellar masses.  Both techniques find that, at fixed M$_{\star}$, AGN hosts with
  log (M$_{\star}$/M$_{\odot}) \gtrsim$ 10.2 are \HI\ rich by a factor of $\sim$ 2.
  However, this gas fraction excess disappears
  when the control sample is additionally matched in star formation rate (SFR), indicating
  that these AGN hosts are actually \HI\ normal. At lower stellar mass,
  the stacking analysis reveals that AGN hosts are \HI\ poor
  at fixed stellar mass.  In the lowest M$_{\star}$ regime probed by our sample,  9 $<$
  log (M$_{\star}$/M$_{\odot}) <$ 9.6, the \HI\ deficit in AGN hosts is a factor of $\sim$ 4,
  and remains at a factor of $\sim$2 even when the control sample is additionally matched
  in SFR.  Our results help reconcile previously conflicting results, by showing that
  matching control samples by more than just stellar mass is critical for a
  rigourous comparison.
  
\end{abstract}


\section{Introduction}

Active galactic nuclei (AGN) have been widely proposed as an effective pathway for
shutting down star formation in galaxies, as they provide an energy mechanism through
which gas can be either heated or removed (see Harrison et al. 2018 for a
recent review).  Indirect observational evidence for the role of AGN in quenching
star formation comes from the connection between the likelihood
that a galaxy is no longer forming stars and the properties of
galaxy centres', such as bulge fraction (Omand et al. 2014), bulge mass
(Bluck et al. 2014; Lang et al. 2014), Sersic index
(Wuyts et al. 2011; Mendel et al. 2013), central stellar mass density
(Cheung et al. 2012; Fang et al. 2013; Woo et al. 2015),
central velocity dispersion (Bluck et al. 2016; Teimoorinia et al.
2016) and black hole mass (Terrazas et al. 2016).  However, it has been
argued that such a `morphology quiescence' relation need not require
a causal connection between an AGN and the cessation of star formation
(Lilly \& Carollo 2016).

Conclusive observational evidence for AGN driven quenching
has remained elusive, with apparently conflicting results in the literature.
For example, although local radio-selected AGN are
hosted by passive elliptical galaxies (e.g. Best \& Heckman 2012), which
could be interpreted as a causal relationship between nuclear activity and
the shut-down of star formation, most AGN are found in star-forming galaxies
(e.g. Hughes \& Cortese 2009; Santini et al. 2012; Rosario et al. 2013).  Indeed,  AGN selected
in the mid-infra red (IR) actually show
elevated star formation rates (Shimizu et al. 2015; Ellison et al. 2016b;
Cowley et al. 2016; Azadi et al. 2017).
Linking AGN to quenching via their star formation rate (SFR) therefore
seems to be highly dependent on the method through which the AGN are
selected.

An alternative approach to linking feedback processes with quenching
is through the study of gas flowing out of the galaxy and outflows are
indeed common in AGN host galaxies (e.g. Mullaney et al. 2013; McElroy et al.
2015; Woo et al. 2016).  However,
there is significant complexity (see e.g. the arguments presented
in Harrison et al 2018) in linking the evident
outflows with the eventual fate of its constituent gas: Although
large masses may be entrained in these outflows, the gas
might not actually escape the galaxy's potential well, and may eventually
become available for star formation once again.  Indeed,
although some studies have proposed that massive outflows could
play an important role in quenching (e.g. Cicone et al. 2014; Baron et al.
2018), other studies have concluded that most AGN-driven winds have velocities
well below the escape value and that low mass fractions are
ejected (e.g. Concas et al. 2017; Davies et al. 2018;
Fluetsch et al. 2018; Roberts-Borsani \& Saintonge 2018).
Even the most powerful local sources
(starbursts and quasars) do not seem adequate to actually remove a significant
fraction of their gas reservoir (e.g. Pereira-Santaella et al. 2018;
Shangguan, Ho \& Xie 2018).  Simulations have similarly concluded
that low redshift galaxies mostly recycle the material in their
outflows, rather than ejecting it (e.g. Muratov et al. 2015; Angles-Alcazar
et al. 2017).  

In this study, we investigate the potential role of AGN in quenching
star formation by assessing the impact of the AGN on the host galaxy's
atomic gas reservoir.  Again, the literature yields conflicting results
on this topic.  Some studies have found that the atomic gas fractions
of AGN hosts are consistent with inactive galaxies (e.g. Fabello
et al. 2011; Gereb et al. 2015), and a similar conclusion has been
drawn from molecular gas fractions (e.g. Saintonge et al. 2012; 2017;
Husemann et al. 2017; Rosario et al. 2018).
However, other studies have found both elevated (e.g. Ho, Darling \&
Greene 2008; Berg et al. 2018) and suppressed (Haan et al. 2008) \HI\
gas reservoirs in and around AGN host galaxies.

There are several subtleties associated with these previous studies
that may contribute to their conflicting conclusions.  For example,
Bradford et al. (2018) have recently suggested that the impact of
AGN feedback could be mass dependent, since they find
evidence of gas depletion only in the low mass galaxies of their sample.
Moreover, several of these previous studies rely on relatively
shallow \HI\ data, which can bias our view of
the `norm' and often necessitates stacking.  The use of deeper data and
accounting for non-detections can significantly change the outcome of
observational comparisons.  For example, the Arecibo
Legacy Fast ALFA (ALFALFA) survey (Giovanelli et al. 2005) is a
large \HI\ survey with relatively shallow depth.
Using ALFALFA, and similar depth targeted observations, Ellison et al. (2015) found that
recently merged galaxies had similar atomic gas fractions to control
galaxies.  However, repeating this experiment with much deeper \HI\
data, and accounting carefully for non-detections, Ellison, Catinella
\& Cortese (2018) have recently shown that post-mergers are actually
a factor of $\sim$ 3 more \HI\ rich than isolated galaxies of the
same stellar mass.

In the work presented here, we use two different approaches to
assess the \HI\ gas fraction of AGN host galaxies.  First, we use
deep, targeted 21~cm measurements from the  Galaxy Evolution Explorer
(GALEX) Arecibo Sloan
Digital Sky Survey (SDSS) Survey (xGASS, Catinella et al. 2018) and
compare \HI\ gas fractions of optically identified AGN host galaxies with a stellar
mass and SFR matched control sample.  However, although xGASS reaches
atomic gas fractions
(\fgas\ = M$_{\rm HI}$/M$_{\star}$) as deep as a few percent, there are only
75 (optically selected) AGN in the xGASS sample.  For a complementary analysis,
we use the large, but shallow, ALFALFA survey (Giovanelli et al. 2005; Haynes et
al. 2018).  Since individual ALFALFA
spectra are an order of magnitude less sensitive than xGASS, we use
an \HI\ spectral stacking approach to measure average atomic gas fractions
for samples of AGN hosts, along with stellar mass and SFR matched control samples.

The paper is laid out as follows.  In Sec. \ref{sec_methods} we describe our 
sample selection and methods.  In Sec. \ref{sec_results}
we present our results for the two methods: gas fractions in individual xGASS
galaxies (Sec. \ref{sec_results_xgass}) and in ALFALFA spectral stacks (Sec.
\ref{sec_results_stacks}).  A discussion is presented in Sec.
\ref{sec_discuss} and a summary in Sec. \ref{sec_summary}.  We adopt a cosmology in which
H$_0$=70 km/s/Mpc, $\Omega_M$=0.3, $\Omega_\Lambda$=0.7.

\section{Methods}\label{sec_methods}

\subsection{Parent galaxy sample and AGN identification}

All of the galaxies used in our analysis are selected from the SDSS data release 7
(DR7).
We make use of the public JHU/MPA catalogs\footnote{https://wwwmpa.mpa-garching.mpg.de/SDSS/}
of stellar masses and emission line fluxes (e.g.
Kauffmann et al. 2003a; Brinchmann 2004).  We will also make use of
SFRs in our analysis.   Although a measurement
of SFR is available in the MPA/JHU catalogs, based on the strength of the
D4000 break (Brinchmann et al. 2004), these can have large uncertainties
for non-star-forming and AGN galaxies (e.g. Rosario et al. 2016).
 We therefore use SFRs obtained from the UV+optical stellar spectral energy
 distribution (SED) fits (no AGN) in the GALEX–-SDSS–-Wide-field Infra-red Survey
 Explorer (WISE) Legacy Catalog
(GSWLC\footnote{http://pages.iu.edu/\textasciitilde salims/gswlc/}; Salim et al. 2016),
using the A2 catalog of Salim, Boquien \& Lee (2018).  Typical
  uncertainties on the stellar masses used in our work are $\sim$ 0.1 dex
  (Kauffmann et al. 2003a).  However, uncertainties on SFRs are highly
  dependent on the SFR itself; for star-forming galaxies uncertainties
  are typically 0.1 dex, but rise to values of $\sim$ 0.6 dex for passive
  galaxies (see Fig. 6 in Salim et al. 2016).

To identify galaxies that host AGN we
first apply the continuum and Balmer S/N cuts described in Scudder et al. (2012),
and then correct emission line fluxes according to a Small Magellanic Cloud dust law as
described by Pei (1992).  We then impose a S/N cut of 5 in the four emission
lines required for the traditional optical AGN classification: [NII], H$\alpha$,
[OIII], H$\beta$, a cut which largely removes galaxies with shock and Low Ionization
Nuclear Emission Region (LINER)-like
spectra.  The diagnostic separation of Kauffmann et al. (2003b) is used to
distinguish AGN from galaxies whose emission spectra are dominated by star
formation\footnote{All of the trends presented in this paper are
  recovered if we use instead the Kewley et al. (2001) AGN selection
  criteria.}.  Specifically, a galaxy is classified as hosting an
  AGN if log([OIII]/H$\beta$) $>$ 0.61/(log([NII]/H$\alpha$) $-$ 0.05) + 1.3 (Eqn. 1 from
Kauffmann et al. 2003b).

\subsection{\HI\ measurements from xGASS}

\begin{figure}
	\includegraphics[width=\columnwidth]{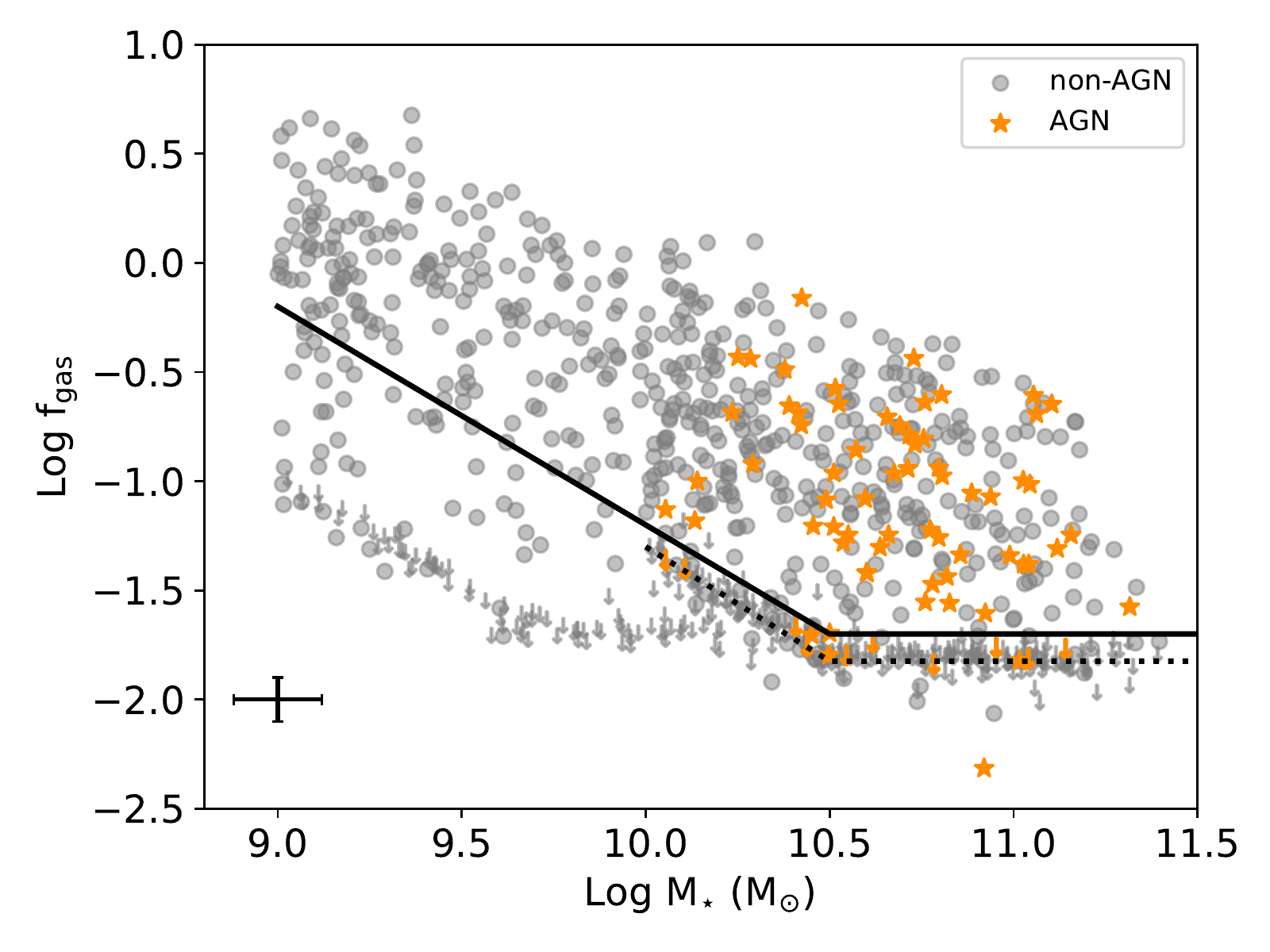}
        \caption{\HI\ gas fractions as a function of stellar mass for AGN
          (orange stars) and non-AGN (grey circles) host galaxies in the xGASS survey.
          For both samples, \HI\ non-detections
          are shown by downward pointing arrows.  
          The dotted line shows the detection threshold of the GASS survey
          (log [M$_{\star}$/ M$_{\odot}] >$ 10), with
          our modified threshold shown by the solid  line that extends to lower
          masses more conservatively
          separates the detections from the upper limits.  The modified detection
          threshold corresponds to  \fgas\ $< $ 2 percent for log (M$_{\star}/M_{\odot}) >$ 10.5
          and log \mhi\ = 8.8  M$_{\odot}$ below that mass. Typical uncertainties
          are shown by the errorbar in the lower left.}
    \label{fgas}
\end{figure}

In the first of the two analyses presented in this work, we compare
the \HI\ gas fractions of individual AGN hosts observed as part of the
xGASS survey (Catinella et al. 2018). The xGASS sample contains $\sim$ 1200 z$<$0.05
galaxies with
measurements of \fgas\ over a mass range 9.0 $<$ log (M$_{\star}/M_{\odot}) <$ 11.5.
\HI\ gas masses are derived in the standard way from 21 cm fluxes, e.g. Eqn 1
  of Catinella et al. (2010), which can be combined with stellar masses to obtain gas
  fractions: \fgas\ = M$_{\rm HI}$/M$_{\star}$.
We use both detections and non-detections in the final xGASS release\footnote{http://xgass.icrar.org/},
but reject any xGASS detections identified in the survey's catalog to have
possible confusion from neighbouring sources. Typical uncertainties
  on \HI\ gas masses in xGASS are 0.05 dex.

In Fig. \ref{fgas} we show the distribution of \HI\ gas fractions 
for the xGASS sample, showing galaxies identified
as hosting AGN as star symbols.  Downward pointing arrows indicate \HI\ upper limits.
There are 75 optically-selected AGN in the xGASS sample, of which 65
are \HI\ detections.  There are 887
non-AGN in the sample (which includes both emission line galaxies not
classified as AGN, as well as galaxies without strong emission lines), of
which 568 are \HI\ detections.  Fig. \ref{fgas} shows the well known bias
that the optically selected AGN in the sample occur at relatively high
masses (e.g. Kauffmann et al. 2003b).
The dotted line in Fig. \ref{fgas} shows the original survey's goal detection
threshold (Catinella et al. 2010).  However, following Ellison et al. (2018),
we make a slightly more conservative demarcation between detections and
non-detections corresponding to  \fgas\ $< $ 2 percent for log
(M$_{\star}/M_{\odot}) >$ 10.5 and log \mhi\ = 8.8  M$_{\odot}$ below that mass
(solid line in Fig. \ref{fgas}).
The exact detection threshold definition does not affect our results.
Detections below the adopted threshold are counted as non-detections for the
statistics in the rest of this paper.

\subsection{Spectral stacking of ALFALFA data}

\begin{figure}
	\includegraphics[width=\columnwidth]{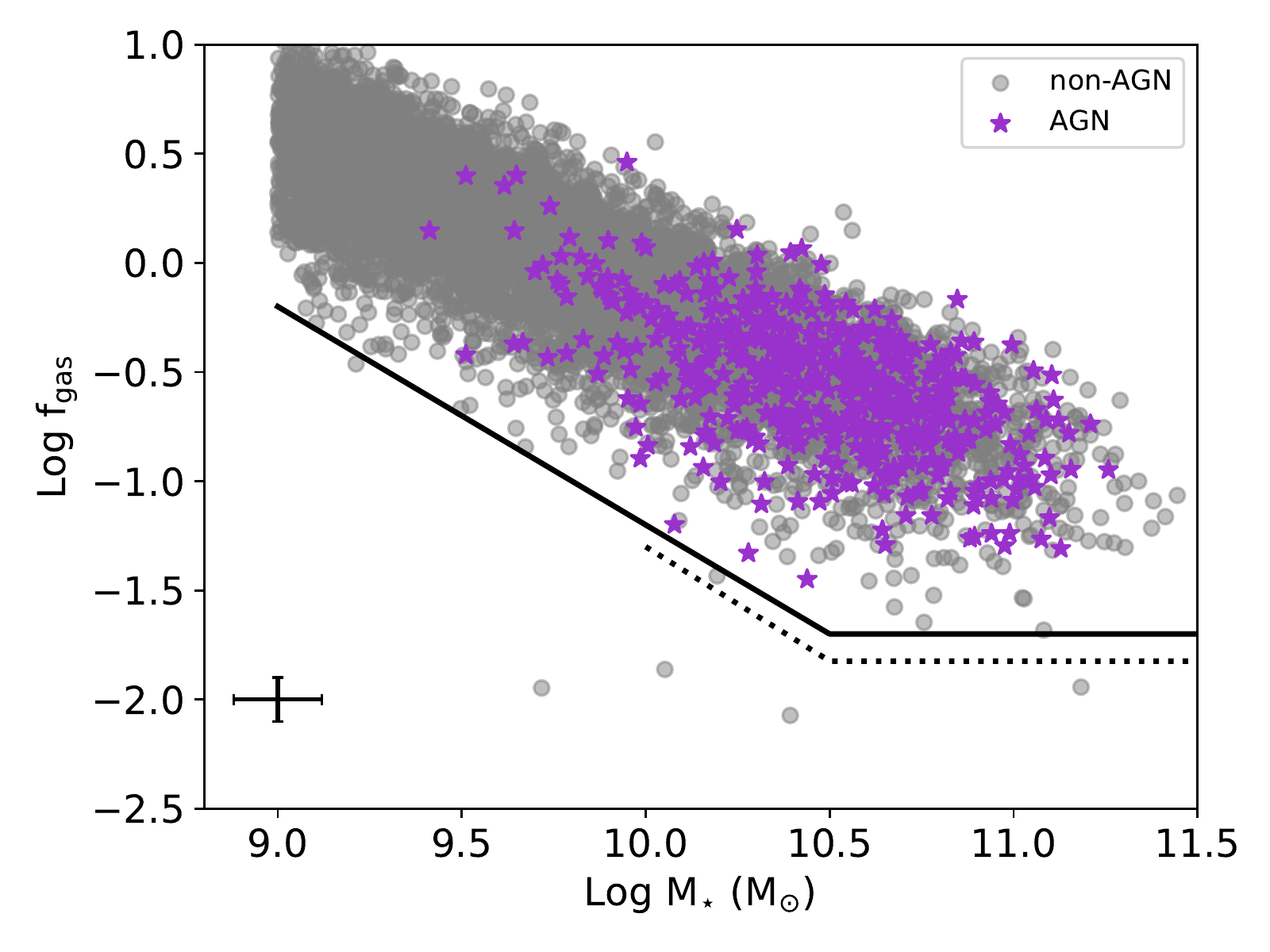}
        \caption{\HI\ gas fractions as a function of stellar mass for AGN
          (purple stars) and non-AGN (grey circles) host galaxies in the ALFALFA sample.
          Only \HI\ detections are available in the ALFALFA catalog.
          The axis ranges and solid and dashed lines (xGASS detection thresholds)
          are the same as shown in
          Fig. \ref{fgas}, for ease of comparison between the two surveys. Typical uncertainties
          are shown by the errorbar in the lower left.} 
    \label{fgas_a100}
\end{figure}

In this work, we use data cubes from the ALFALFA 100 percent sample (Haynes et al. 2018).
The final data release of the ALFALFA blind-HI survey is the largest census
of atomic gas content in the local Universe to date, mapping over 7000 square degrees of
sky out to a redshift of $z\sim0.06$.

The subsamples of galaxies used for our stacking analysis (Sec \ref{sec_results_stacks})
are drawn from a parent sample of 34,142 galaxies that is selected according to stellar
mass (9 $\le$ log (M$_{\star}$/M$_{\odot}$) $\le$ 11.5) and redshift (0.02 $\le z \le$ 0.05)
from the overlap in volume between the full ALFALFA survey and SDSS DR7. The increase in
the number of galaxies between the parent sample used in this work and that of Brown et al.
(2015) comes from our use of the recently available ALFALFA 100 percent datacubes.

After requiring that
galaxies in the ALFALFA parent sample are also included in the GSWLC, and
that there is an AGN (according to Kauffmann et al 2003b)
classification available, our sample is reduced to 28,678 galaxies, which
we refer to hereafter as `the ALFALFA sample'.
From this reduced sample, we identify 1562 optically-selected AGN,
509 of which are detected by ALFALFA. There are thus 27,116 non-AGN host galaxies
in the ALFALFA sample that can be used as a comparison sample, of which 7722 have \HI\ detections.

In Fig. \ref{fgas_a100} we present the distribution of
gas fractions for the individual 21~cm detections in the ALFALFA sample, taken
directly from the public ALFALFA catalog.  In contrast to xGASS (Fig. \ref{fgas}),
no upper limit information is provided in the ALFALFA catalogs.
The distribution
of points in Fig. \ref{fgas_a100} therefore shows the ALFALFA sample's
distribution of gas fractions for \HI\ detections, but does not represent the full set of data included
in our spectral stacks (which include non-detections as well, see below).
As in Fig. \ref{fgas}, we plot AGN host galaxies as stars and non-AGN hosts
as circles\footnote{Throughout
  this paper, we facilitate the visual comparison of the xGASS and ALFALFA analyses
  by using orange and purple symbols in figures that refer to these two samples, respectively.}.
We set the x- and y-axis
ranges of Fig. \ref{fgas_a100} (ALFALFA) to match that of Fig, \ref{fgas} (xGASS),
and also reproduce
the xGASS detection thresholds in the figure, to facilitate comparison between the
two surveys and demonstrate the relative depth of xGASS compared with ALFALFA.
Fig. \ref{fgas_a100} shows that ALFALFA's \HI\ detections are typically
  0.5 dex shallower than the xGASS detection threshold at fixed stellar mass.

The ALFALFA analysis presented here does not use gas fractions from the public
catalogs, but rather produces spectral stacks directly from the original data cubes.
Following the steps outlined in Brown et al. (2015), we extract individual \HI\
spectra for each galaxy using a 4$\times$4 arcminute aperture and 5500 \kms\
velocity cut centered on its position in the ALFALFA datacubes. 
The \HI\ stacking is performed by
co-adding 21 cm spectra that are aligned in velocity (redshift) to yield an average atomic gas
measurement for a given sample of galaxies regardless of whether they are individually detected
in \HI. The stochastic nature of each individual \HI\ spectrum's noise means that the
signal-to-noise ratio of the stacked spectrum increases as function of the square root of the
galaxies in each stack. Note that, following Brown et al. (2015), we weight the individual
spectra in each stack by the galaxy's stellar mass and luminosity distance so that the
final stacked spectrum is in 'gas fraction units'. For a complete description of our
data processing and stacking methodology, we refer the reader to Brown et al. (2015).

\section{Results}\label{sec_results}

\subsection{Gas fractions of AGN in the xGASS sample}\label{sec_results_xgass}

\begin{figure}
	\includegraphics[width=\columnwidth]{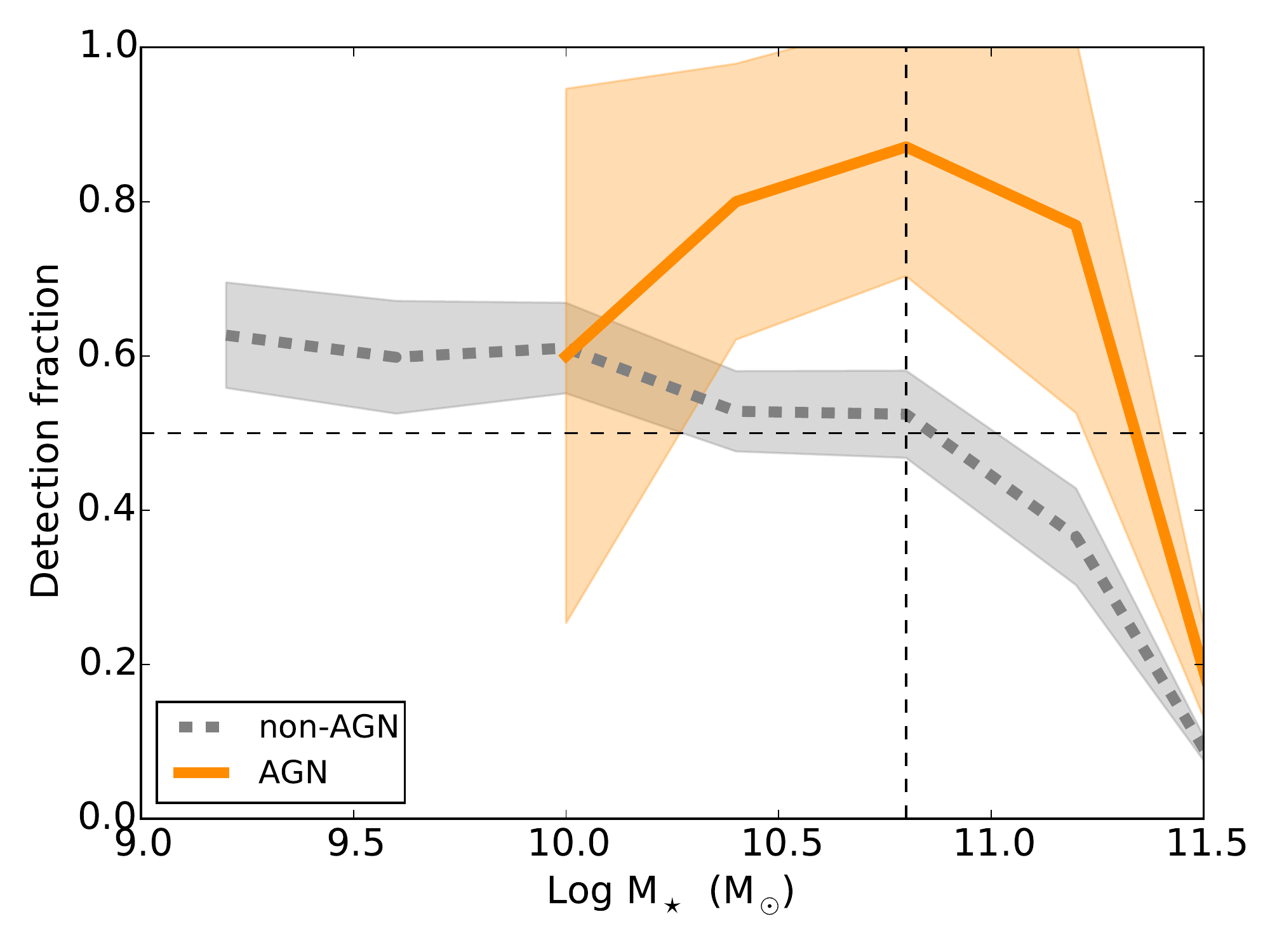}
        \caption{The \HI\ detection fraction of AGN (solid orange line)
          and non-AGN (dashed grey line) host galaxies
          in the xGASS sample with 1$\sigma$ bounds shaded.  The vertical
          dashed line corresponds to log M$_{\star} =$ 10.8 M$_{\odot}$
          above which the xGASS detection fraction drops below 50 per cent
        (horizontal dashed line).}
    \label{det_frac}
\end{figure}

In Fig. \ref{det_frac} we show the fraction of \HI\ detections in the xGASS sample as a
function of stellar mass, with 1$\sigma$ bounds shown as the shaded
regions.  The AGN host \HI\ detection fraction (shown in solid orange) is significantly
above the non-AGN host \HI\ detection fraction (shown in dashed grey) for most
stellar masses.  The higher detection fraction amongst the AGN host
galaxies suggests that their gas fractions are systematically higher
than their non-active counterparts at fixed stellar mass.

For a quantitative assessment of the difference in \fgas\ between AGN
and non-AGN hosts, we follow Ellison et al. (2018),
and compute the \HI\ gas fraction offset, \dfgas.
\dfgas\ is computed for each AGN host galaxy in turn by identifying
all non-AGN galaxies in the xGASS sample of the same stellar mass (within $\pm$
0.15 dex) and computing the difference between log \fgas\ in the
AGN galaxy and median value of the matched control galaxies.  
There are typically between 100 -- 200 non-AGN control galaxies
matched to each AGN host.

The detection fraction in the non-AGN sample exceeds 50 per cent (horizontal
dashed line in Fig. \ref{det_frac}) for stellar
masses log (M$_{\star}$/M$_{\odot}) \le$ 10.8 (vertical dashed line in Fig. \ref{det_frac}).
Therefore, the median gas fraction and \dfgas\ are
well constrained in this mass regime, even accounting for non-detections.
Of the 75 AGN in the full xGASS sample (Fig \ref{fgas}), 50 have
log (M$_{\star}$/M$_{\odot}) \le$  10.8 and hence have reliable \dfgas\
determinations.  Of these, 41 are \HI\ detections (and are above the adopted
detection threshold shown by the solid line in Fig. \ref{fgas}) and nine
are non-detections (or detections below the solid line in Fig. \ref{fgas}).

\begin{figure}
	\includegraphics[width=\columnwidth]{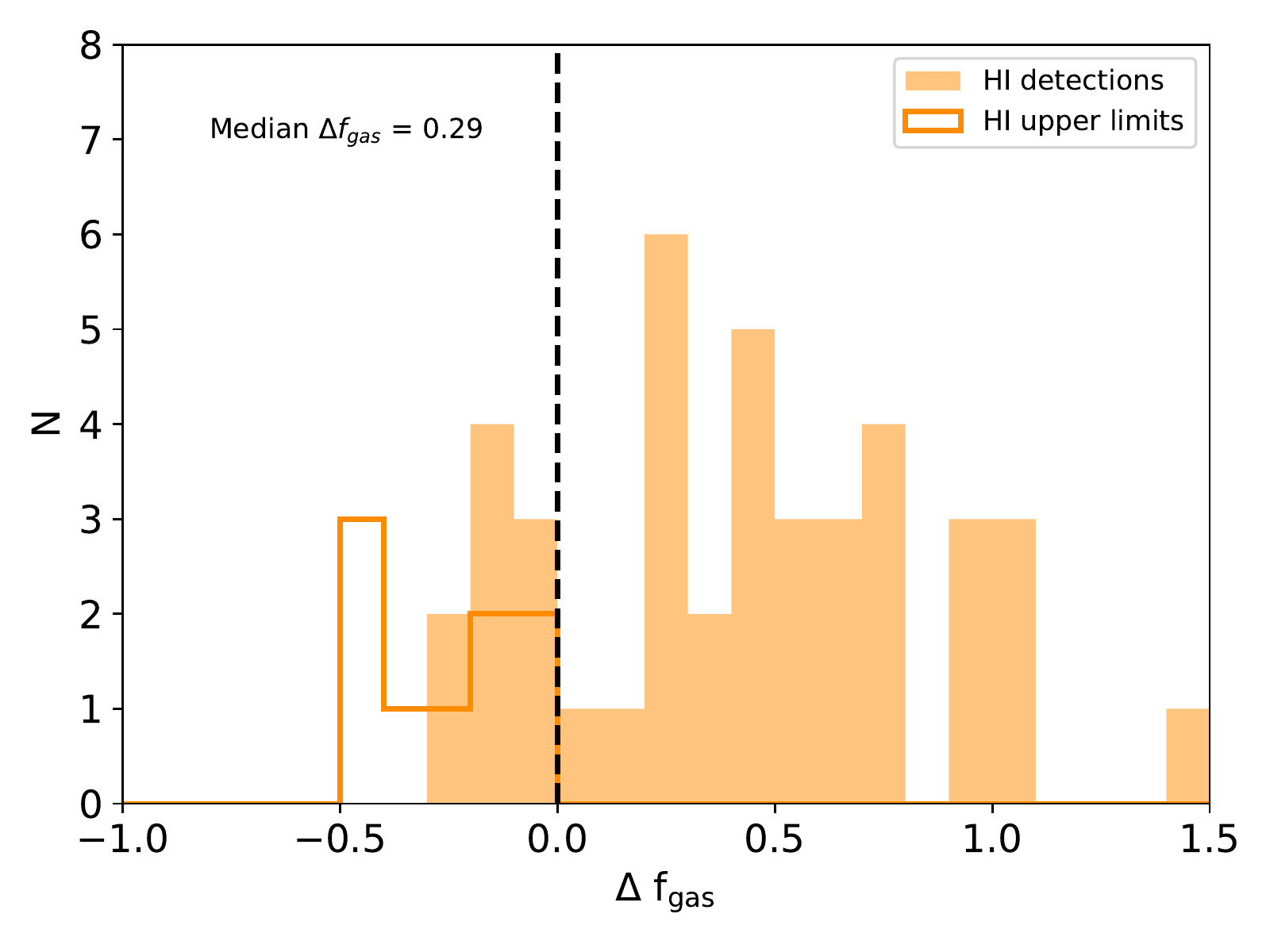}
        \caption{\dfgas\ distribution for AGN host galaxies, matched to
          control galaxies in M$_{\star}$,
          in the xGASS sample with log (M$_{\star}$/M$_{\odot}) \le$ 10.8. Both
          detections and upper limits are considered in the control sample.
          The solid/open histograms show \dfgas\ for \HI\ detections/upper limits
          respectively.}
    \label{dfgas}
\end{figure}

\begin{figure}
	\includegraphics[width=\columnwidth]{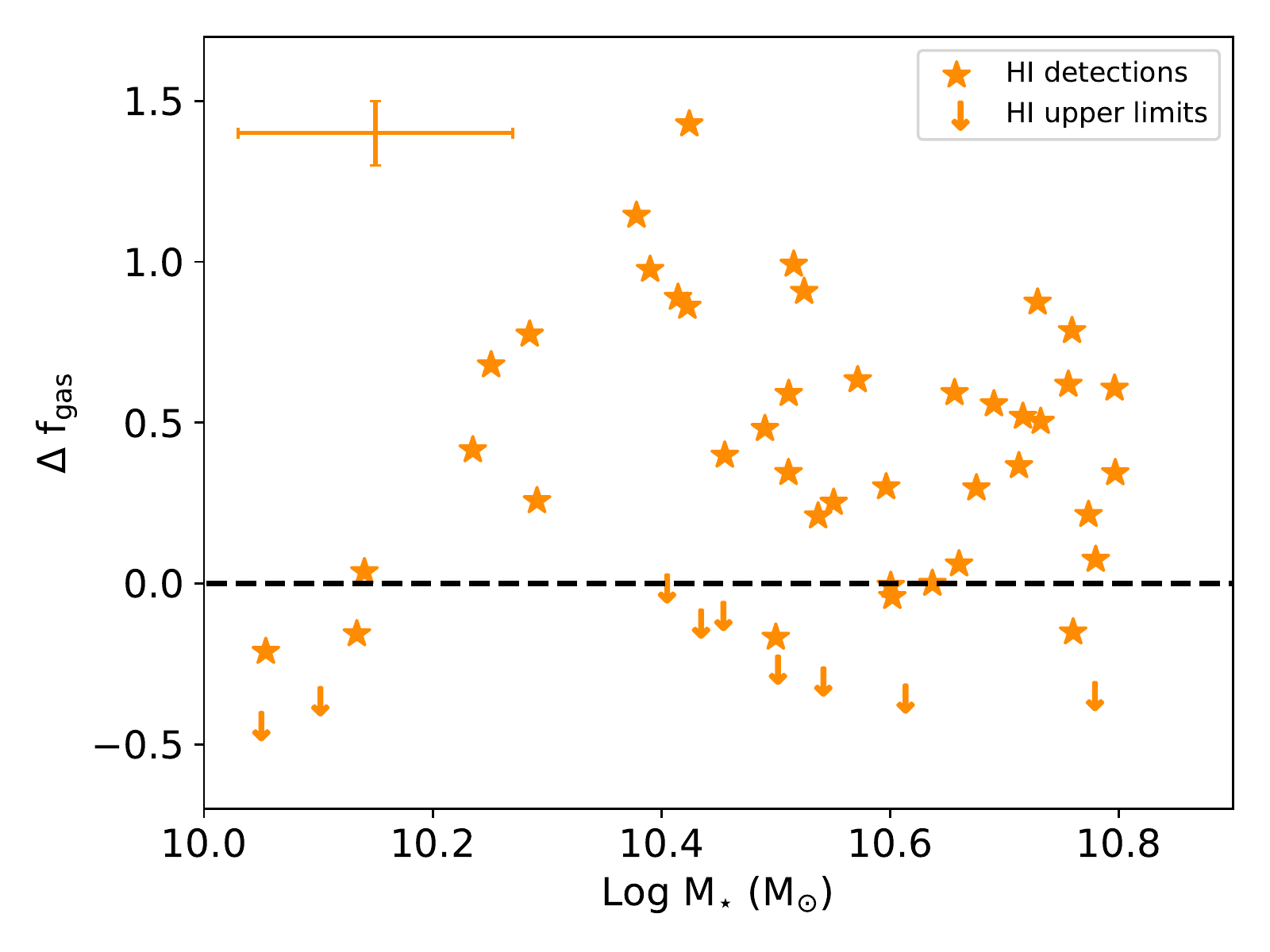}
        \caption{\dfgas\ as a function of stellar mass for AGN hosts in the xGASS sample.  Solid stars
          indicate \HI\ detections and downward pointing arrows are upper
        limits. Typical uncertainties
          are shown by the errorbar in the upper left.}
    \label{dfgas_mass}
\end{figure}

Fig. \ref{dfgas} shows the distribution of \dfgas\ for the sample of
50 AGN hosts with log (M$_{\star}$/M$_{\odot}) \le$  10.8.  The \HI\ non-detections
lead to upper limits in the calculation of \dfgas; these nine galaxies
are shown as the open histogram in Fig. \ref{dfgas}.  The median \dfgas\
of the full sample of 50 AGN hosts (including the nine upper limits) is $+0.29$
dex, representing a factor of two \fgas\ enhancement on average at fixed
stellar mass. 
In Fig.  \ref{dfgas_mass} we plot \dfgas\  versus total stellar mass.
The enhanced \fgas\ is fairly consistent across the full stellar
mass range of our sample, although, intriguingly, the five
lowest mass optical AGN hosts, with 10.0 $<$ log (M$_{\star}$/M$_{\odot}) <$  10.2,
do not show enhancements.  We re-visit the mass dependence of \dfgas\
in the next Section, where the ALFALFA stacking analysis offers better
statistics at lower stellar masses.

Although optically selected AGN have slightly lower SFRs than other main
sequence galaxies (e.g. Ellison et al. 2016b; Leslie et al. 2016), they are nonetheless
usually hosted by star-forming and `green valley' galaxies, with a
relative paucity amongst passive galaxies (e.g.  Hughes \& Cortese 2009;
Santini et al. 2012; Rosario et al. 2013).  
The elevated gas fractions in AGN hosts at fixed stellar mass may be
a manifestation of this tendency to avoid the passive population, whereas non-AGN
control galaxies have a wide range of SFR at a given stellar mass.

\begin{figure}
	\includegraphics[width=\columnwidth]{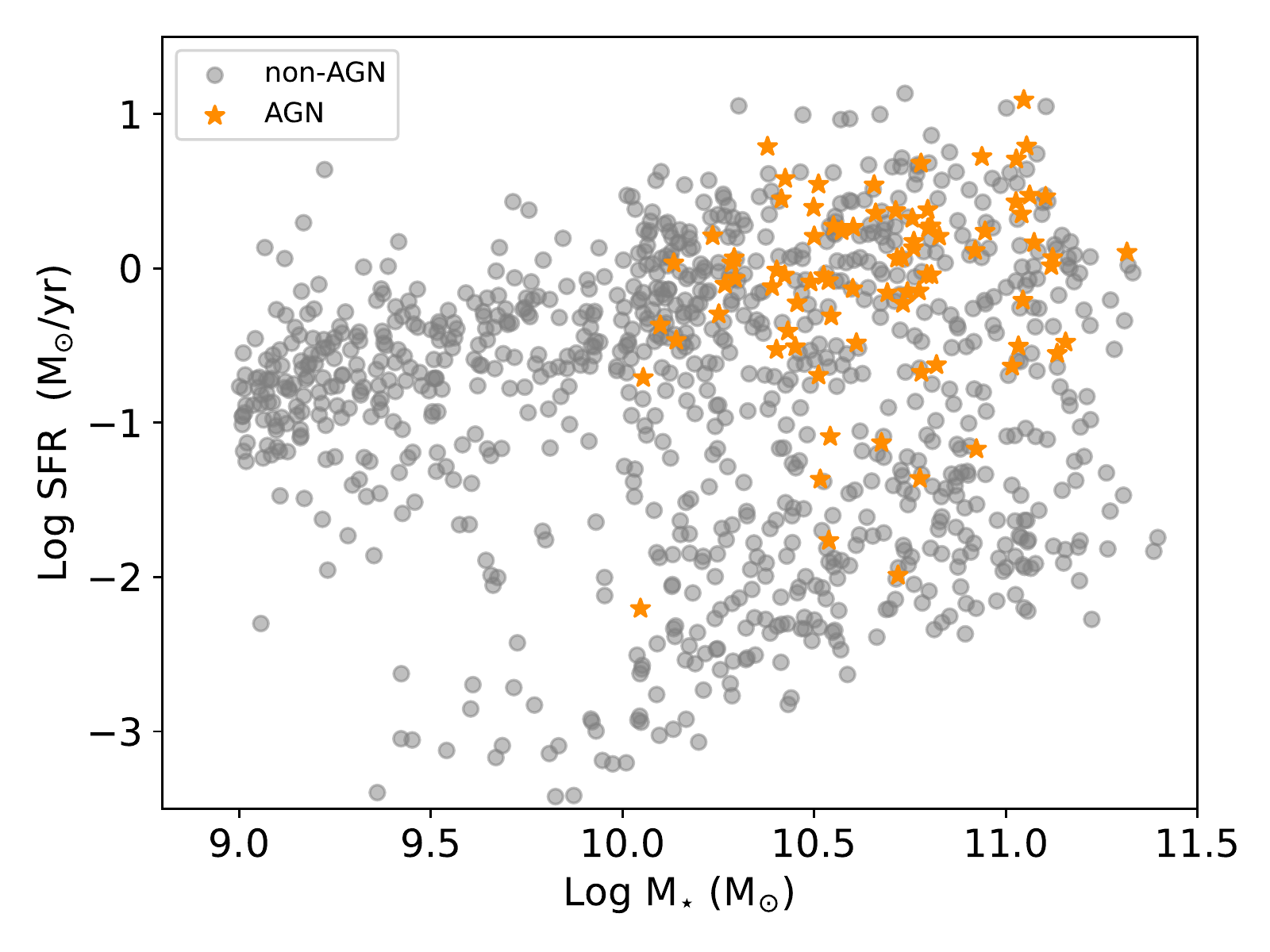}
        \caption{Distribution of SFR versus stellar mass for the xGASS sample. Non-AGN
          hosts are plotted with grey circles and AGN hosts in orange stars.  AGN hosts tend to
          lie on the star forming main sequence, or green valley, with relatively few
        quenched hosts.}
    \label{xgass_MS}
\end{figure}

In Fig. \ref{xgass_MS} we investigate
the tendency of AGN to be hosted in star forming galaxies in the xGASS sample
by plotting the distribution of SFR vs. stellar mass for the xGASS non-AGN 
sample with grey circles and AGN hosts in xGASS in orange stars.  
Fig. \ref{xgass_MS} shows that, as expected, the AGN hosts in the xGASS sample
lie predominantly along the star-forming main sequence, with relatively few located in
either the green valley or in the regime of `quenched' galaxies.
We therefore repeat our
calculation of \dfgas, but now match in both M$_{\star}$
and SFR (equivalent to matching in sSFR at fixed mass).
The requirement of a measured SFR in the GSWLC
slightly reduces the size of the AGN sample, to 39 \HI\ detections
and 9 \HI\ non-detections (down from 41 and 9 in the original sample).
We require that the SFRs of control galaxies be matched to that
of the AGN to within $\pm$ 0.1 dex.
The additional requirement of a SFR match greatly reduces the
number of controls for each AGN host, from more than 100 when
only M$_{\star}$ is matched to typically 10--20 with matched
SFR and  M$_{\star}$.  

\begin{figure}
	\includegraphics[width=\columnwidth]{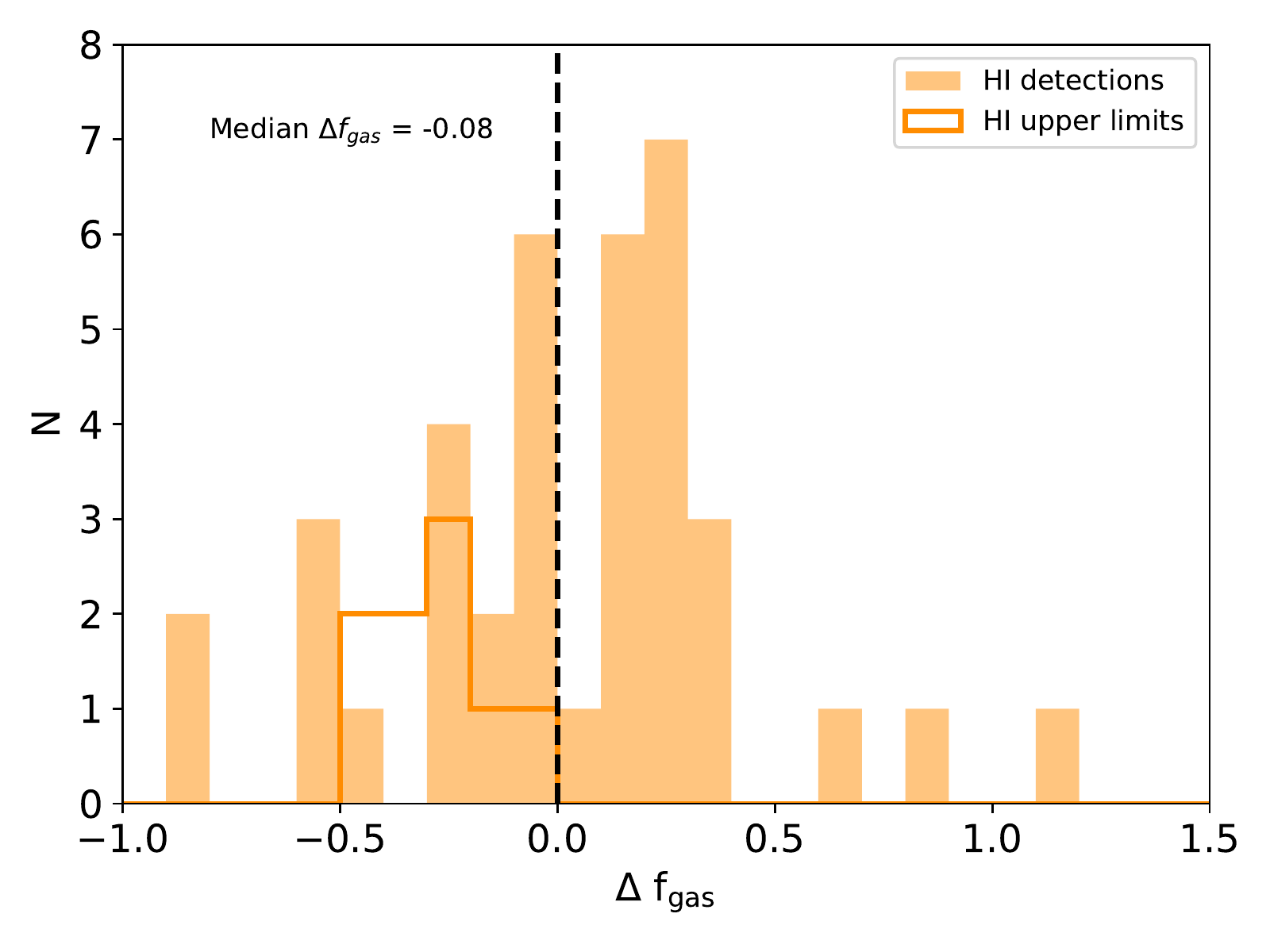}
        \caption{\dfgas\ distribution for AGN host galaxies matched in
            both M$_{\star}$ and SFR for
          the xGASS sample with log (M$_{\star}$/M$_{\odot}) \le$ 10.8  considering both
          detections and upper limits in the control sample.  SFRs are taken
          from the GSWLC of Salim et al. (2016).
          The solid/open histograms show \dfgas\ for \HI\ detections/upper limits respectively.}
    \label{dfgas2}
\end{figure}

In Figure \ref{dfgas2} we show the distribution of \dfgas\
when SFR is included with M$_{\star}$ in the matching criteria.
In contrast to the mostly positive values in Figure \ref{dfgas},
Figure \ref{dfgas2} shows that the inclusion of SFR in the
matching process now yields a distribution
of \dfgas\ that is more symmetric around zero.  The median (including non-detections)
is slightly negative ($-0.08$ dex) but there is no statistically
significant deviation from a gaussian distribution centred at zero.

We repeat the SFR matching analysis using SFRs from two other catalogs.
The first alternative is using the SFRs published in
the xGASS catalog by Janowiecki et al. (2017),
who use GALEX and WISE photometry.
Second, we determine SFRs from the total IR luminosities (\LIR)
obtained by the artificial
neural network calibration of the Herschel Stripe 82 overlap (Ellison et al.
2016a) as a proxy for SFR, requiring $\sigma_{ANN} \le$ 0.1.  We find
qualitatively consistent distributions of \dfgas\ when using any of these three
(Salim et al. 2018; Janowiecki et al. 2017; Ellison et al. 2016a)
SFR indicators.  In all cases, we find that once matched in both SFR
and stellar mass, the AGN host gas fractions are consistent with those
of the non-AGN control galaxies.

The results from this section show that comparing the AGN to a non-AGN control
sample matched in stellar mass alone will yield mis-leading results.  The gas fraction
enhancement seen in Figs. \ref{dfgas} and \ref{dfgas_mass} are purely the result
of AGN preferentially being hosted in star forming galaxies.  Once taking this
`bias' into account, there is no statistically significant difference between
the \HI\ gas fractions of AGN and non-AGN host galaxies.  

\subsection{Gas fractions of AGN in the ALFALFA stacks}\label{sec_results_stacks}

\begin{figure}
	\includegraphics[width=\columnwidth]{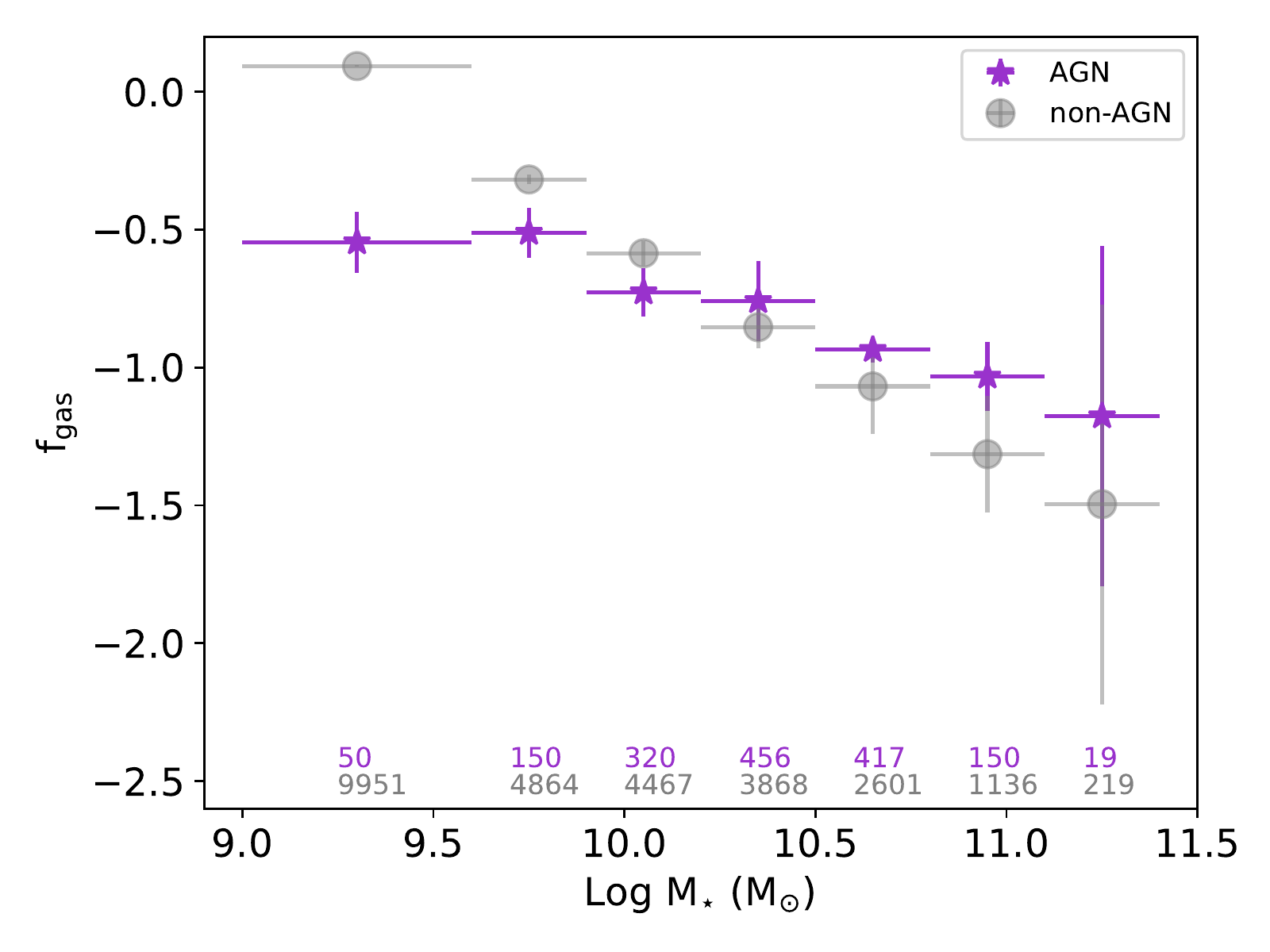}
	\includegraphics[width=\columnwidth]{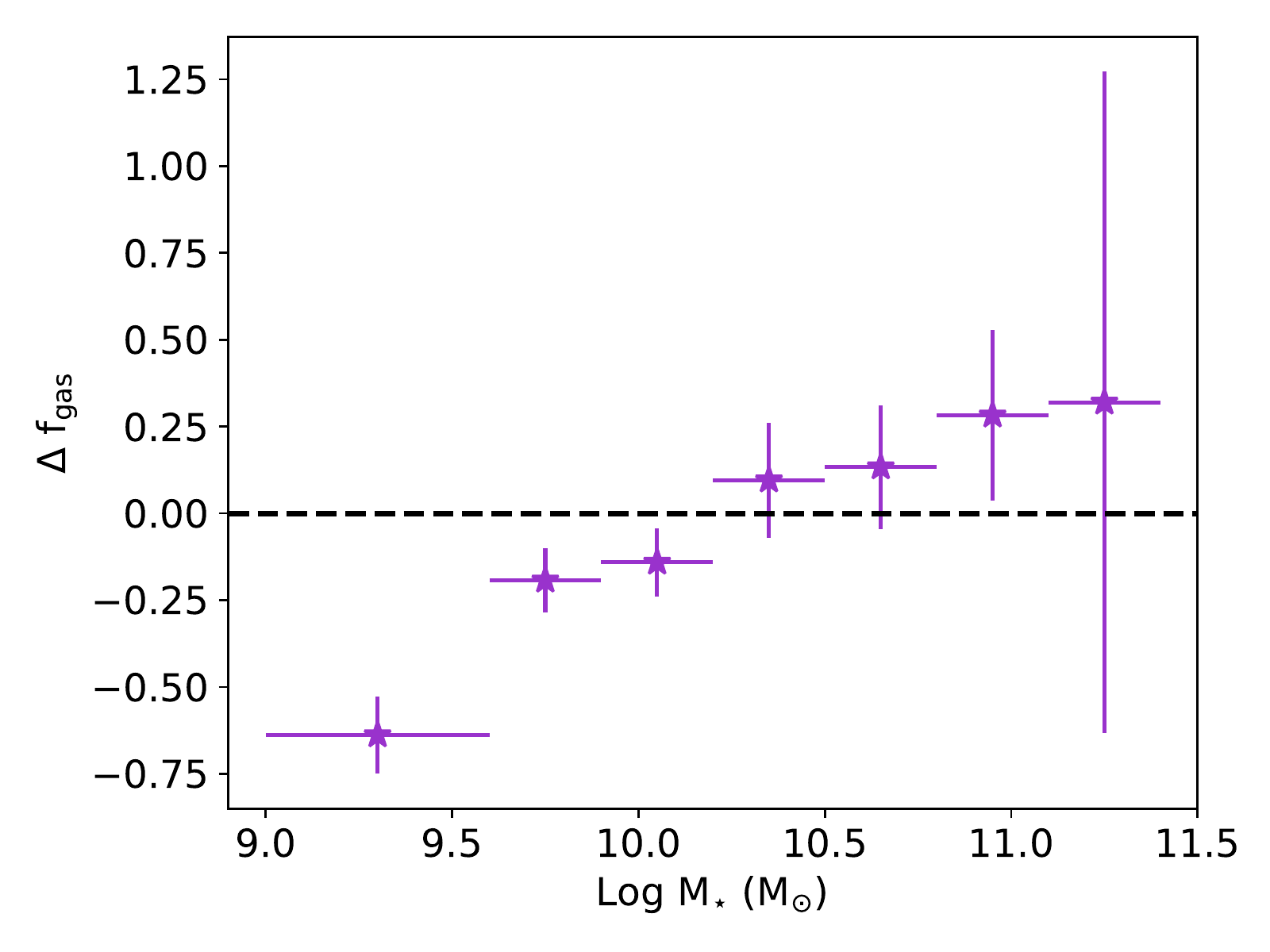}
        \caption{Top panel: Measured \fgas\ for the AGN (purple stars) and non-AGN
          (grey circles) hosts in the ALFALFA spectral stacks in bins of stellar mass. Numbers
          under each data point indicate how many individual spectra contributed
          to each stack.  Bottom panel:
          \dfgas, computed as the difference between the AGN and non-AGN host stack gas fractions
          in the upper panel. For both panels, each data point is plotted at the centre of its
          mass bin (which is close to the mean mass in that bin), with the x-axis error bars
          showing the width of the bin in which the stack was constructed.}
    \label{dfgas_a100}
\end{figure}

We now turn to the ALFALFA spectral stacking analysis.  
In the top panel of Fig. \ref{dfgas_a100} we show the \HI\ gas fractions as a function of
stellar mass for the AGN (star symbols) and non-AGN (circles) host ALFALFA spectral stacks.
The numbers underneath the data points indicate how many individual spectra contributed
to a given stack.  
The stellar mass range for each spectral stack is 0.3 dex, i.e. a mass
  bin width that is equivalent to
the stellar mass matching criterion of $\pm$ 0.15 dex used in the xGASS
matching procedure.  The exception is in the lowest stellar mass bin where
a paucity of AGN necessitates a broader stellar mass range of 0.6 dex in order
to achieve sufficient numbers and sensitivity in the stack.
Each data point in Fig \ref{dfgas_a100} is plotted at the centre of its mass bin (which
is close to the mean mass in that bin), with the x-axis error bars showing the
width of the bin in which the stack was constructed.
In the lower panel of Fig. \ref{dfgas_a100}, we compute a gas fraction offset analogous
to that computed for the individual xGASS galaxies. However, whereas in the xGASS analysis
we were able to compute a gas fraction offset on a galaxy-by-galaxy basis, for the ALFALFA
spectral stacks, we compute \dfgas\ as simply the difference of log \fgas\ for the AGN and
non-AGN host stacks (i.e, the difference between star and circle symbols in the upper panel of Fig
\ref{dfgas_a100}).

The top panel of Fig. \ref{dfgas_a100} shows that at stellar masses
log (M$_{\star}$/M$_{\odot}$) $>$ 10.2 AGN hosts are more gas-rich than the control
in given stellar mass bin.
The bottom panel of the figure shows that the gas fraction enhancement
is typically $\sim$ 0.25 dex at these high masses.  Although the error bars in
Fig. \ref{dfgas_a100} are quite large, the trend of elevated gas
fractions at high stellar mass is systematic.  Increasing the size of the
stellar mass bins yields a more statistically significant result, although
for consistency with the bins used in the xGASS analysis (which uses a
  mass matching tolerance of 0.15 dex) we have kept the
original $\pm$0.15 dex (i.e. 0.3 dex width) mass bins.
The results at high stellar mass in Fig. \ref{dfgas_a100} are broadly consistent with our finding of
elevated \fgas\ in individual xGASS AGN host galaxies, e.g. Fig. \ref{dfgas_mass}
which shows gas fraction enhancements above log (M$_{\star}$/M$_{\odot}$) $\sim$ 10.2
and with a sample median enhancement of \dfgas = 0.29 dex (Fig \ref{dfgas}).

The larger size of the ALFALFA sample, compared to xGASS, means that we include
AGN host galaxies down to lower stellar masses.  Fig. \ref{dfgas_a100} reveals that at
log (M$_{\star}$/M$_{\odot}$) $<$ 10, a regime not probed by the xGASS sample,
the gas fractions of the AGN hosts are now lower than in the stellar-mass
matched control sample.  This is a statistically significant result,
particularly for the lowest mass AGN host galaxies in our sample, 9 $<$
log (M$_{\star}$/M$_{\odot}$) $<$ 9.6 (a total of 50 AGN), which are a factor $\sim$ 4
more \HI\ poor than non-AGN hosts of the same stellar mass.

\begin{figure}
	\includegraphics[width=\columnwidth]{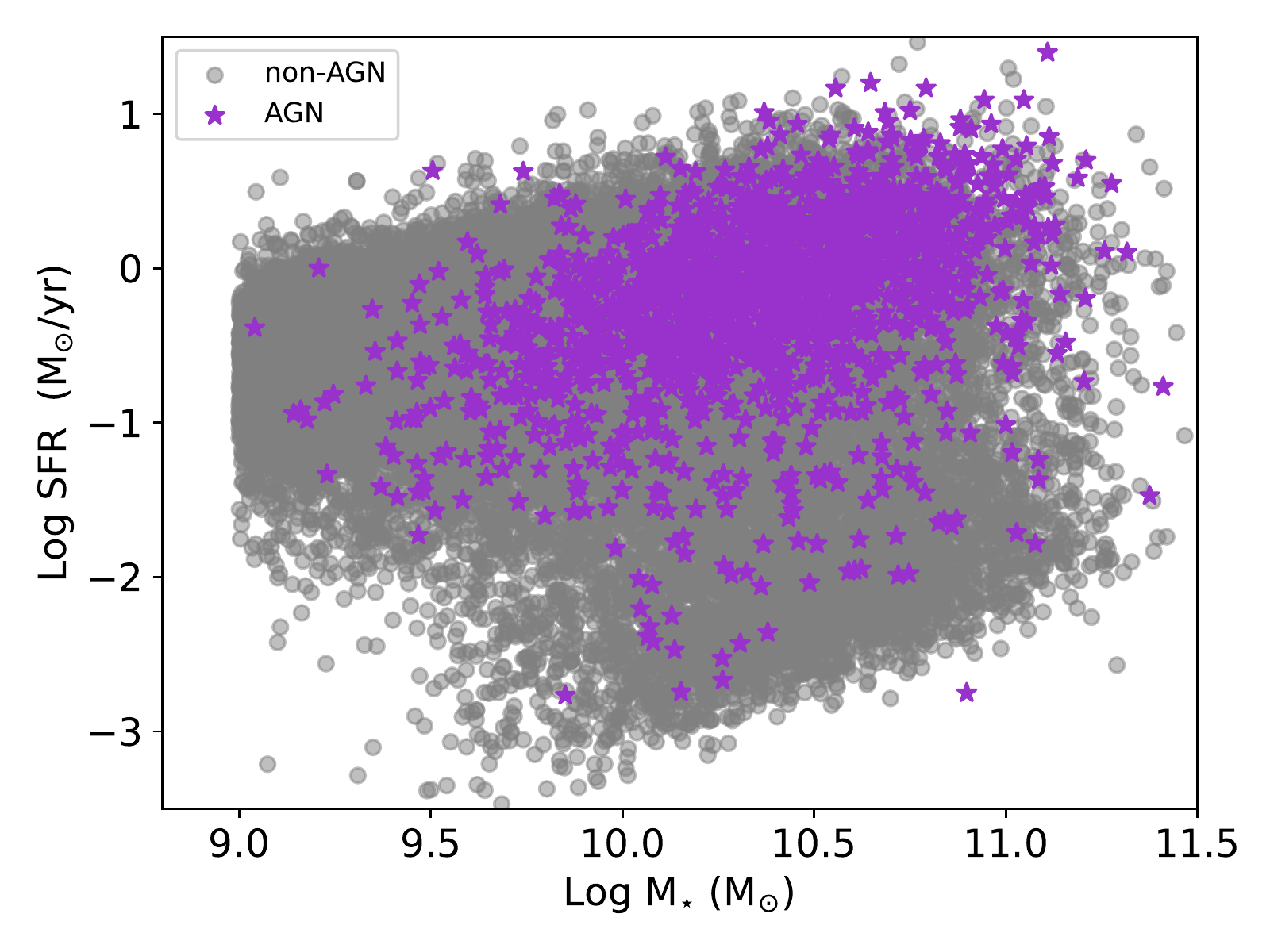}
        \caption{Distribution of SFR versus stellar mass for the ALFALFA sample. Non-AGN
          hosts are plotted with grey circles and AGN hosts in purple stars.  AGN hosts tend to
          lie on the star forming main sequence or green valley with relatively few
        quenched hosts.}
    \label{A100_MS}
\end{figure}

As for the xGASS analysis, we next investigate the additional impact of matching
the control sample not only in stellar mass, but also in SFR.  The distribution of
SFRs and stellar masses for the ALFALFA sample is shown in Fig. \ref{A100_MS}.
As was the case in the xGASS sample (Fig. \ref{xgass_MS}), AGN hosts in the
ALFALFA sample are also primarily located on the star-forming main sequence
or in the green valley, motivating the need to match in SFR.

\begin{figure}
	\includegraphics[width=\columnwidth]{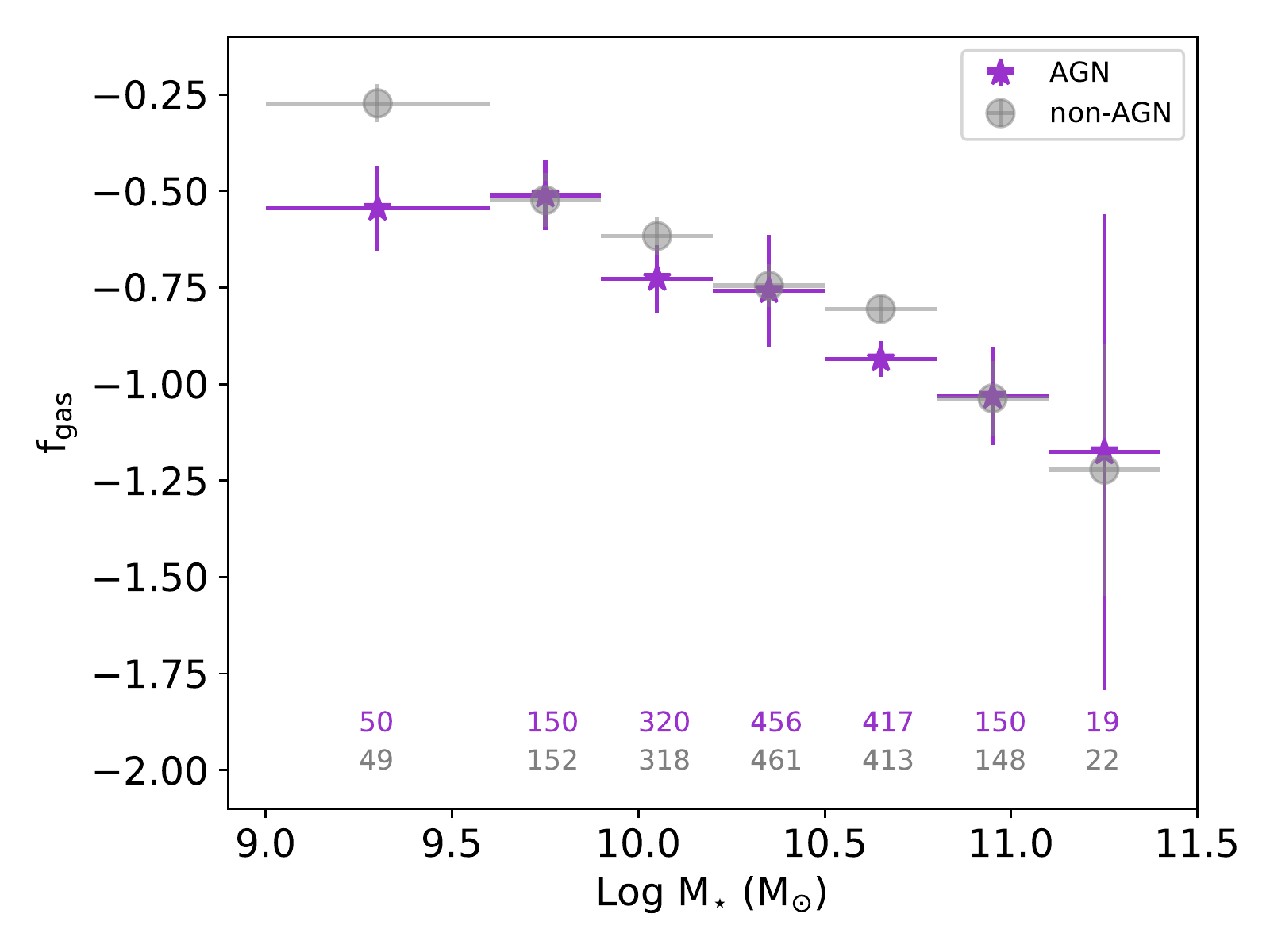}
	\includegraphics[width=\columnwidth]{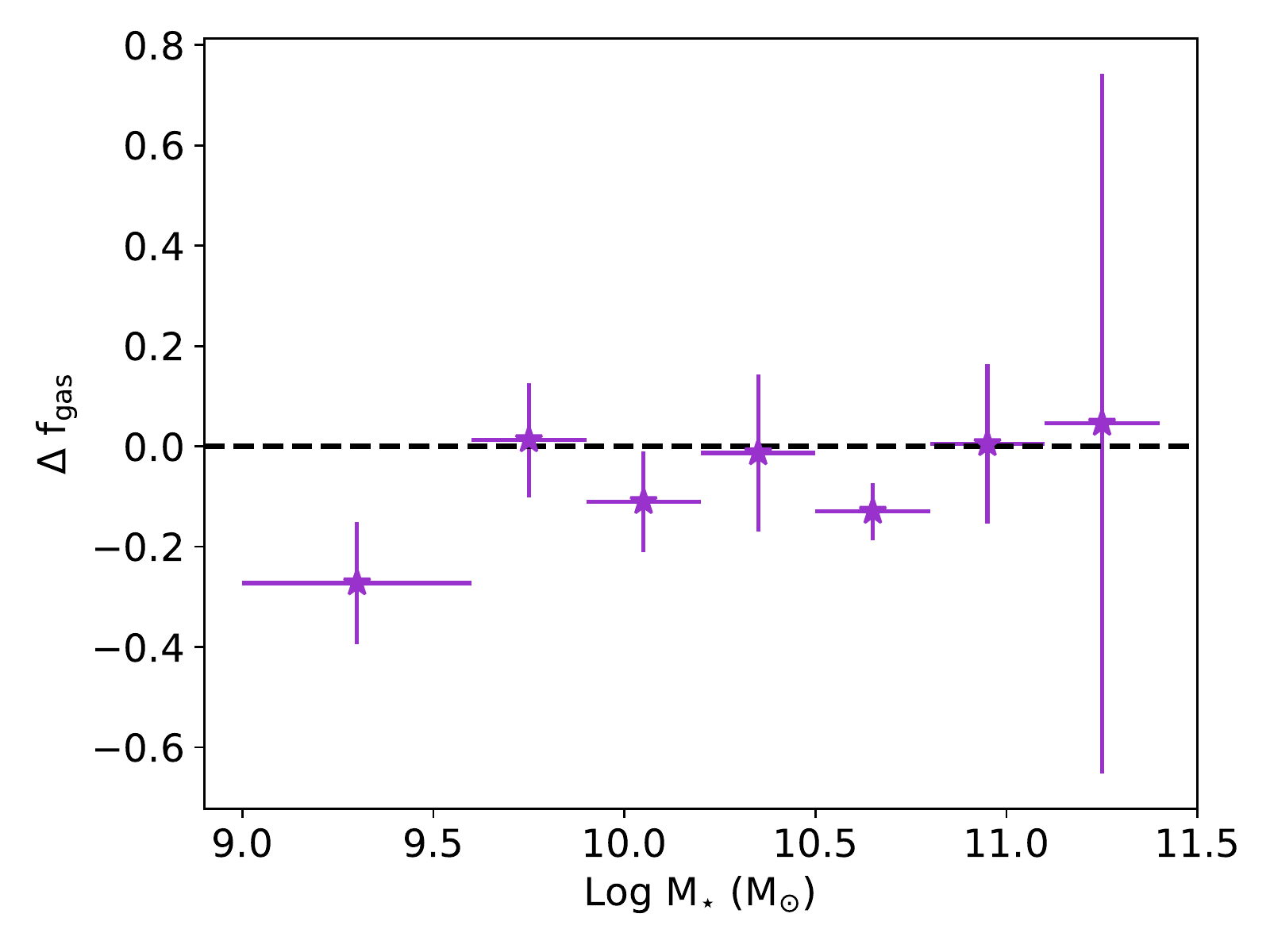}
        \caption{Top panel: Measured \fgas\ for the AGN (purple stars) and non-AGN
          (grey circles) hosts in the ALFALFA spectral stacks in bins of stellar mass with control
          matching in both M$_{\star}$ and SFR.  Numbers
          under each data point indicate how many individual spectra contributed
          to each stack.  Bottom panel:
          \dfgas, computed as the difference between the AGN and non-AGN host stack gas fractions
          in the upper panel.  For both panels, each data point is plotted at the centre of its
          mass bin (which is close to the mean mass in that bin), with the x-axis error bars
          showing the width of the bin in which the stack was constructed.}
    \label{dfgas_sfr_a100}
\end{figure}

Again,  the ALFALFA spectral
stacking requires a slightly different practical procedure than for the
xGASS sample, in which individual AGN hosts are matched to controls.  For every AGN galaxy
that is included in a given stellar mass stack, we identify the non-AGN control galaxy
in ALFALFA that is the closest simultaneous match in both M$_{\star}$ and
SFR (again, SFRs are taken from the A2 catalog of Salim et al. 2018).
Due to the large number of non-AGN galaxies in ALFALFA (27,116), the control
sample can be very tightly matched in these quantities, with both the SFRs and
M$_{\star}$ values typically matched to within 0.02 dex or less (which is about
an order of magnitude less than the uncertainty in these values) of the AGN value.
Nonetheless, due to the combination of well defined mass bins used in our spectral stacking
(nominally spanning 0.3 dex), and the permitted tolerance of the SFR matching,
the number of individual spectra (numbers beneath the data points in Fig.
\ref{dfgas_sfr_a100}) that go into the AGN and non-AGN host stacks can
differ by a few.

Fig. \ref{dfgas_sfr_a100} is analogous to the data shown in Fig. \ref{dfgas_a100},
but now has the additional SFR matching included in the stacked data.  The upper panel
shows the \fgas\ values in stacks binned by stellar mass for the AGN hosts (purple stars) and
controls (grey circles).  The lower panel of Fig. \ref{dfgas_sfr_a100} shows
\dfgas\ -- the difference between the AGN hosts and control gas fractions. Fig. \ref{dfgas_sfr_a100}
shows that there is
no longer any indication for an \HI\ excess at log (M$_{\star}$/M$_{\odot}$) $>$ 10.2,
or in any other stellar mass regime.  Therefore, as we previously concluded based
on the xGASS analysis, the \HI\ excess at fixed stellar mass is entirely a
result of AGN hosts being preferentially located on the star-forming main sequence.
Once this is accounted for by matching in SFR, both the xGASS and ALFALFA
stacking analyses find the majority of AGN host galaxies to be \HI\ normal.
The possible exception to \HI\ normalcy in AGN hosts remains in the lowest stellar mass bin probed by
the ALFALFA stacking analysis. In the 9 $<$ log (M$_{\star}$/M$_{\odot}$) $<$ 9.6
stellar mass bin in the lower panel of Fig. \ref{dfgas_sfr_a100} AGN hosts
still appear to be \HI\ poor by a factor of two (down from a factor of four
at fixed stellar mass without SFR matching).

\section{Discussion}\label{sec_discuss}

\subsection{Comparison to previous results - a reconciliation}

Previous comparisons of the \HI\ content of AGN host galaxies that have used
stacked 21 cm spectra have found no difference in average \fgas\
compared with a non-AGN control sample (Fabello et al. 2011; Gereb et
al. 2015).  Likewise, using the relatively shallow ALFALFA
survey to define `normal' \HI\ gas fractions, Bradford et al.
(2018) have found no difference in atomic gas fractions in most AGN hosts.
However, Bradford et al. (2018) report a possible deficit in \fgas\
in AGN hosts for galaxies with log (M$_{\star}$/M$_{\odot}$) $< 9.5$.
\HI\ deficits in AGN host galaxies were also reported by Haan et al. (2008).
Conversely, one of the earliest investigation of global \HI\
gas fractions of AGN hosts, by Ho et al. (2008) reported the
perhaps surprising result that AGN actually host higher
gas fractions for their morphological type.  Using the complementary
technique of quasar absorption line spectroscopy to probe the circumgalactic
medium, 
Berg et al. (2018) have also recently reported a factor
of three excess \lya\ absorption in sightlines through AGN hosts.

The literature has therefore variously
reported normal, elevated and reduced \HI\ gas fractions in and around AGN
host galaxies. Different
techniques, survey depths, assessment of comparison samples and treatment
of limits likely contributes to the different conclusions drawn
by previous works.

We have re-visited the quantification of \HI\ gas fractions in
optically selected AGN host galaxies in the SDSS using two complementary approaches:
individual \fgas\ measurements for a sample of 75 optically selected AGN
for which we have \HI\ measurements measured down to detection thresholds of
only a few per cent, and spectral stacking for 1562 AGN in the ALFALFA survey.
Both techniques take non-detections into account and use matched
control samples for comparison drawn from the same datasets.
Our experiment is therefore both well controlled for systematics,
and uses two different techniques for cross-validation.

Both the individual xGASS measurements and the ALFALFA stacks show
that AGN hosts with log (M$_{\star}$/M$_{\odot}$) $\gtrsim 10.2$ have \HI\ gas fractions
that are a factor of $\sim$ 2 higher than non-AGN of the same stellar mass (Figs
\ref{dfgas} and \ref{dfgas_a100}).
Although the xGASS sample only contains AGN with log (M$_{\star}$/M$_{\odot}$) $> 10$,
the ALFALFA stacks allow us to probe down to
log (M$_{\star}$/M$_{\odot}$) $\sim$ 9.0. For AGN hosts in the ALFALFA sample with
log (M$_{\star}$/M$_{\odot}$) $< 10$, we find a significant deficit
of \HI, at fixed stellar mass.  This deficit is greatest in the lowest
stellar mass bin, 9.0 $<$ log (M$_{\star}$/M$_{\odot}$) $< 9.6$, where the \HI\
gas fraction is lower than the non-AGN hosts by a factor of $\sim$4 (Fig \ref{dfgas_a100}).
The two complementary techniques used here therefore yield consistent results at
high stellar masses (where the samples overlap and are comparable) and show that
the gas fractions of AGN hosts are enhanced for their stellar mass.

However, a comparison between AGN and non-AGN hosts at fixed stellar mass
fails to take into account that the former tend to inhabit star-forming
or green valley galaxies, whereas the latter show a broad range of SFRs,
including passive galaxies.  We have shown that when this `bias'
is accounted for, by matching the non-AGN control sample in both stellar mass
and SFR, the enhanced \fgas\ in AGN hosts disappears
(Figs. \ref{dfgas2} and \ref{dfgas_sfr_a100}).  This is confirmed
with both the xGASS and ALFALFA analyses.  Therefore, conclusions
concerning the gas fraction in AGN hosts depend not only on the stellar
mass of the galaxy, but also on the experimental design and control
sample parameters. 

Our results help resolve several apparent inconsistencies that
have been previously reported in the literature.  For example,
Ho et al. (2008) and Berg et al. (2018) have both reported
\HI\ excesses in AGN hosts.  Both of these studies focused on
relatively massive galaxies and matched either directly (Berg
et al. 2018) or indirectly (using luminosity, Ho et al. 2008)
on galaxy stellar mass.  The results of both Ho et al. (2008)
and Berg et al. (2018) are therefore in agreement with the factor
of two \HI\ excess for AGN host galaxies with log (M$_{\star}$/M$_{\odot}) \gtrsim$
10.2 that we find in xGASS and ALFALFA (Figs \ref{dfgas} and \ref{dfgas_a100}).

On the other hand, Fabello et al. (2011) and Gereb et al. (2015)
found that AGN hosts are \HI\ normal.  Both of these studies take
into account the $NUV - r$ colour of the AGN hosts and compare their
\HI\ gas fractions to controls of similar colours.  This approach
is similar to our comparison of \HI\ gas fractions that are matched
in both stellar mass \textit{and} SFR.  Therefore, our finding that
matching in both of these parameters results in gas fractions that
are consistent between AGN hosts and controls  (Figs \ref{dfgas2} and
\ref{dfgas_sfr_a100}) is in good agreement with Fabello et al. (2011)
and Gereb et al. (2015).

Finally, Bradford et al. (2018) found that AGN host galaxies with
log (M$_{\star}$/M$_{\odot}) <$ 9.5 are \HI\ poor.  Again, this
is consistent with our findings from the ALFALFA stacks, both
at fixed M$_{\star}$ (where AGN hosts are a factor of 4 more \HI-poor than
the controls: Fig \ref{dfgas_a100}), and also with additional SFR matching
(factor of 2: Fig \ref{dfgas_sfr_a100}).

\subsection{Dependence of gas fraction on [OIII] luminosity}

\begin{figure}
	\includegraphics[width=\columnwidth]{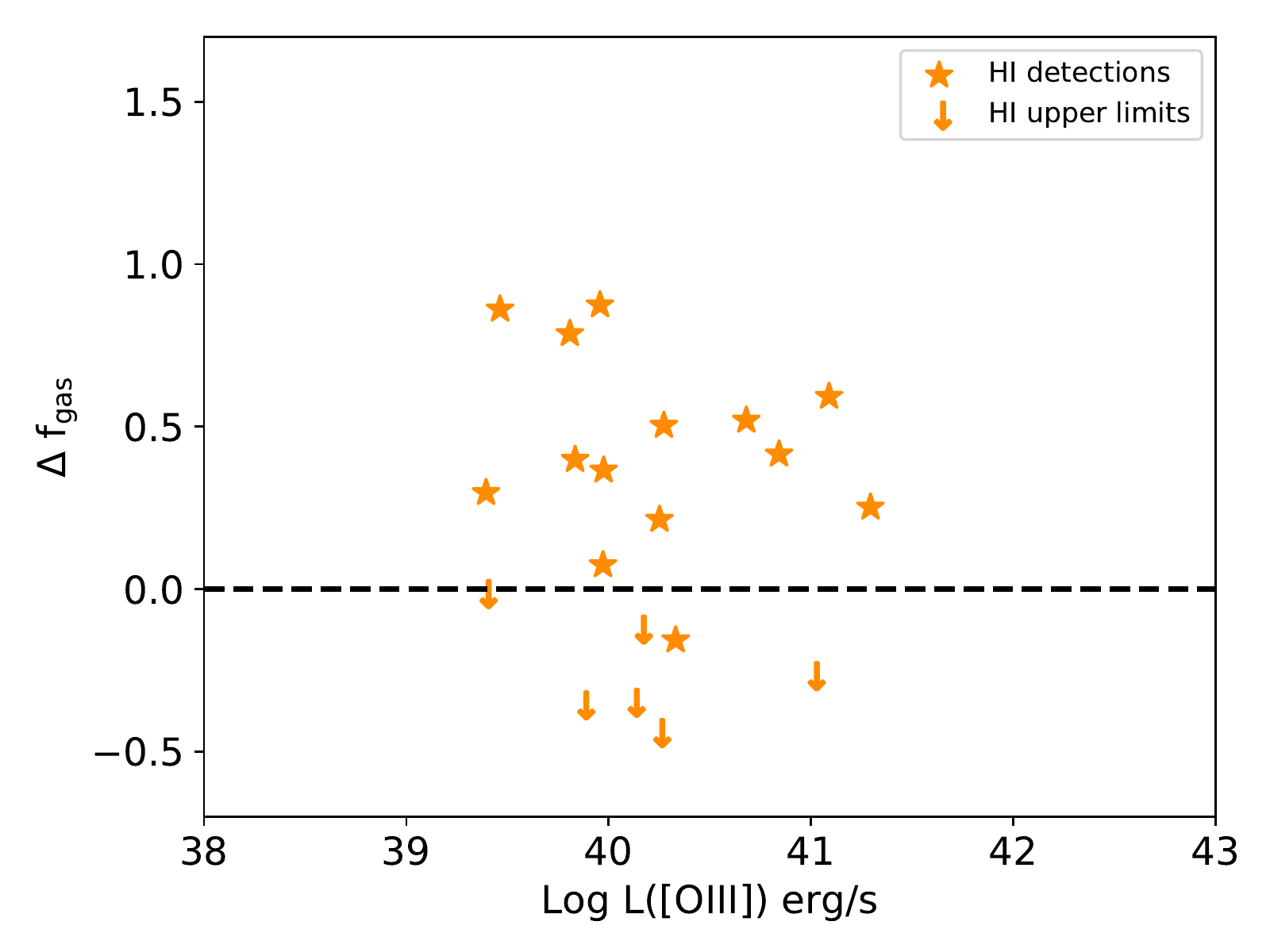}
	\includegraphics[width=\columnwidth]{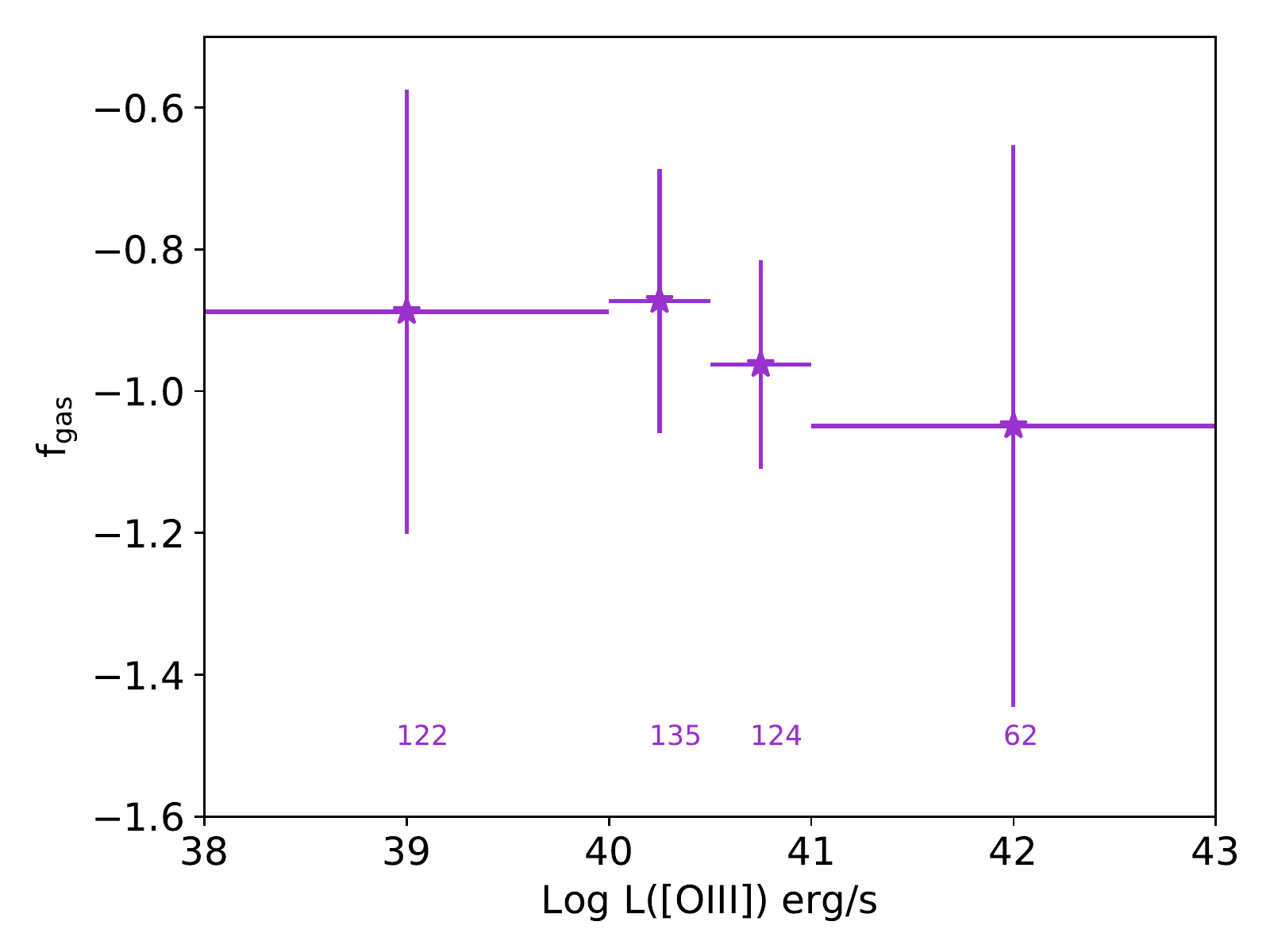}
        \caption{Top panel: \dfgas\ as a function of
          [OIII] line luminosity for AGN hosts in the xGASS sample. Lower panel:  \fgas\
          as a function of [OIII] line luminosity for the ALFALFA stacks.  [OIII]
          luminosities are only computed for galaxies classified as AGN
          by the Kewley et al. (2001) criterion, and have no bolometric correction
          applied. Bottom panel: \fgas\ from ALFALFA stacks in bins of 
          [OIII] line luminosity. Numbers
          under each data point indicate how many individual spectra contributed
          to each stack.  Both panels are plotted on over the same
        range of [OIII] luminosity for ease of comparison.}
    \label{dfgas_lo3}
\end{figure}

For galaxies whose emission lines are dominated by photo-ionization from
the AGN, the [OIII] line luminosity can be used as
an indicator of AGN luminosity (e.g. Kauffmann et al. 2003b).
Contamination from star formation can be minimized by selecting AGN
according to the Kewley et al. (2001) criteria.
In the upper panel of Fig \ref{dfgas_lo3} we plot \dfgas\ versus [OIII] luminosity for
individual galaxies in the xGASS
sample identified as AGN using the Kewley et al. (2001) cut.
We have not attempted to make a bolometric correction to
the [OIII] luminosities as this represents a simple (but unknown) multiplicative
factor for the x-axis, which does not impact any correlation.
We find no trend between the gas fraction enhancement and [OIII] luminosity.

In the lower panel of Fig \ref{dfgas_lo3} we plot \HI\ gas fraction
versus [OIII] luminosity for ALFALFA spectral stacks that are now constructed
in bins of L(O[III]).  Again, these stacks are made only for AGN identified
as AGN using the Kewley et al. (2001) criteria.
Since L(O[III]) is not a relevant (in the sense of measuring nuclear accretion)
quantity for non-AGN
galaxies, there is no corresponding control stack, and therefore we can only
look for variation in \fgas\ and not \dfgas\ in the spectral stacks.
Nonetheless, the results for the ALFALFA sample again indicate that there
is no dependence of gas fraction on the AGN luminosity.  Therefore, neither
the xGASS, nor the ALFALFA results provide any evidence that the gas reservoir
is systematically more affected by higher AGN luminosities.  In a complementray work,
Shangguan et al. (2018) have recently quantified the
total gas fractions of low redshift quasars, whose luminosities
are considerably greater than our Seyfert sample, and again found
that the majority of the hosts retain high gas fractions.

\section{Summary and conclusions}\label{sec_summary}

AGN feedback has been proposed to lead to the cessation of star formation
via the heating and/or removal of the galactic gas reservoir.  However,
previous measurements of the \HI\ gas fraction in optically selected AGN
hosts have found values largely consistent with the non-AGN population
in massive galaxies
(e.g. Fabello et al. 2011; Gereb et al. 2015; Bradford et al. 2018).
Conversely, other works have found that AGN hosts may be in fact be
relatively \HI-rich (Ho et al. 2008; Berg et al. 2018).

In order to understand these apparently conflicting results, we re-visit the
assessment of atomic gas
fractions in AGN hosts using data from two complementary surveys, and
using two different methods of analysis.  First, we use
a sample of 75 optically selected AGN  from the relatively deep xGASS survey,
and show that, at fixed stellar mass, the \HI\ detection fraction is higher
for AGN than non-AGN hosts.
With a careful accounting for non-detections, we can compute gas fraction
`offsets' compared to a mass matched non-AGN sample for 50 of the AGN hosts
with 10 $<$ log (M$_{\star}$/M$_{\odot}$) $<$ 10.8.  We find that AGN host
\HI\ gas fractions are elevated by a factor of two compared to non-AGN
galaxies of the same stellar mass. However,  we suggest that this effect
is driven by the tendency of optical AGN to inhabit star forming
host galaxies.  When we additionally
match our non-AGN sample in SFR, the median gas fraction of the AGN hosts is
consistent with the controls.  We conclude that in the
regime 10 $<$ log (M$_{\star}$/M$_{\odot}$) $<$ 10.8 the gas fractions
of optically selected AGN are consistent with non-AGN.

The second dataset studied in this work is a sample of 1562 optically
selected AGN from the 100 per cent ALFALFA survey.  Although the ALFALFA
survey is relatively shallow, spectral stacking permits an effective way
to quantify gas fractions in sub-samples of the data and can take into
account non-detections in individual spectra.  We show that,
in agreement with the xGASS results, \HI\ gas fractions in AGN hosts are elevated
by a factor of $\sim$ 2 for log (M$_{\star}$/M$_{\odot}$) $\gtrsim$ 10.2
at fixed stellar mass, compared with the non-AGN control sample.
Also in agreement with the xGASS analysis, this excess disappears when
the stacks are additionally matched in SFR.  The ALFALFA stacking analysis
therefore supports our conclusion from the xGASS analysis, that once
properly matched in both M$_{\star}$ and SFR, optically selected AGN
in massive host galaxies are \HI\ normal.

The larger size of the ALFALFA
sample permits us to extend our analysis to lower stellar masses than
is possible with the xGASS sample.  We find that, at fixed stellar mass
(with no SFR matching), AGN hosts with log (M$_{\star}$/M$_{\odot}$) $<$ 10.0
are \HI\ poor.  For the lowest stellar masses, 9.0 $<$ log (M$_{\star}$/M$_{\odot}$)
$<$ 9.6, this paucity is about a factor of 4, and persists even when the
SFR matching is implemented, albeit with lower magnitude (factor of two).
The low gas fractions at low stellar mass are qualitatively consistent with
the recent results of Bradford et al. (2018) and are consistent with reduction
(either through removal or heating)
of the atomic gas reservoir by AGN in the dwarf regime.

Our results help to reconcile apparently conflicting results in the literature,
by showing that conclusions concerning the relative \HI\ content of AGN hosts
depends critically on how the control sample is constructed.
In the context of AGN feedback and gas removal, we only find evidence of this
process at the lowest stellar masses of our sample (log (M$_{\star}$/M$_{\odot}$) $<$ 10
at fixed stellar mass and 9 $<$ log (M$_{\star}$/M$_{\odot}$)
$<$ 9.6 for fixed stellar mass and SFR).  Moreover, neither the xGASS nor the
ALFALFA AGN samples show any dependence of
their gas fraction properties on AGN luminosity. Taken together,
our results indicate that widespread removal of
the host galaxy reservoir by
the AGN is not a significant process for the majority of galaxies in the low
redshift universe.

\section*{Acknowledgements}

SLE gratefully acknowledges support from an NSERC Discovery Grant.
Parts of this research were conducted by the Australian Research Council Centre of 
Excellence for All Sky Astrophysics in 3 Dimensions (ASTRO 3D), through project number 
CE170100013.  We are grateful to Samir Salim for proving SDSS objIDs for
the GSWLC and to Martha Haynes for providing access to the ALFALFA data cubes.

The xGASS and ALFALFA surveys, on which this work is based, were
conducted at the
Arecibo Observatory.
The Arecibo Observatory is operated by SRI International under a
cooperative agreement with the National Science Foundation
(AST-1100968), and in alliance with Ana G. M{\'e}ndez-Universidad
Metropolitana, and the Universities Space Research Association.

Funding for the SDSS and SDSS-II has been provided by the Alfred P. Sloan Foundation, the Participating Institutions, the National Science Foundation, the U.S. Department of Energy, the National Aeronautics and Space Administration, the Japanese Monbukagakusho, the Max Planck Society, and the Higher Education Funding Council for England. The SDSS Web Site is http://www.sdss.org/.

The SDSS is managed by the Astrophysical Research Consortium for the Participating Institutions. The Participating Institutions are the American Museum of Natural History, Astrophysical Institute Potsdam, University of Basel, University of Cambridge, Case Western Reserve University, University of Chicago, Drexel University, Fermilab, the Institute for Advanced Study, the Japan Participation Group, Johns Hopkins University, the Joint Institute for Nuclear Astrophysics, the Kavli Institute for Particle Astrophysics and Cosmology, the Korean Scientist Group, the Chinese Academy of Sciences (LAMOST), Los Alamos National Laboratory, the Max-Planck-Institute for Astronomy (MPIA), the Max-Planck-Institute for Astrophysics (MPA), New Mexico State University, Ohio State University, University of Pittsburgh, University of Portsmouth, Princeton University, the United States Naval Observatory, and the University of Washington.

\end{document}